\pdfoutput=1
\documentclass[journal,twoside,web]{ieeecolor}
\usepackage{generic}
\usepackage{cite}
\usepackage{amsmath,amssymb,amsfonts}
\usepackage{graphicx}
\usepackage{textcomp}
\usepackage{array}
\usepackage{multirow}
\usepackage{url}

\usepackage{booktabs}

\def\BibTeX{{\rm B\kern-.05em{\sc i\kern-.025em b}\kern-.08em
    T\kern-.1667em\lower.7ex\hbox{E}\kern-.125emX}}

\newcommand{\kt}{k_t^2}

\newcolumntype{M}[1]{>{\centering\arraybackslash}m{#1}}

\makeatletter
\def\leftmark{\ifnum\value{page}>1 \rightmark\fi}
\makeatother

\begin{document}

\bstctlcite{IEEEexample:BSTcontrol}

\raggedbottom

\title{FR3-Band Capacitive-Piezoelectric Thin-Film LN Resonator for Heterogeneous Integration}

\author{Vakhtang Chulukhadze, Jiajun Huang, Marco Galli, Tzu-Hsuan Hsu, Udit Rawat, Yao Yu, \\Siddhartha Ghosh, Bichoy Bahr, and Ruochen Lu
\thanks{This work was supported by the National Science Foundation (NSF) under CAREER Award No.\ 2339731. (Corresponding author: Vakhtang Chulukhadze.) V. Chulukhadze, J. Huang, T.-H. Hsu, and R. Lu are with the Chandra Family Department of Electrical and Computer Engineering, The University of Texas at Austin, Austin, TX 78712 USA (e-mail: vatoc@utexas.edu). M. Galli and S. Ghosh are with the Department of Electrical and Computer Engineering, Northeastern University, Boston, MA 02115 USA. U. Rawat, Y. Yu, and B. Bahr are with Kilby Labs, Texas Instruments, Dallas, TX 75243 USA.}}

\IEEEaftertitletext{\vspace{-2\baselineskip}}
\maketitle

\begin{abstract}
This letter presents a capacitively coupled piezoelectric lateral-field-excited bulk acoustic wave resonator (XBAR) in single-crystal lithium niobate (LN) for frequency range 3 (FR3, 7.125-24.25 GHz). We develop a platform that simultaneously maintains high electromechanical efficiency (\(k_t^2\)) and quality factor (\(Q\)) while alleviating challenges associated with parasitic galvanic interconnections and is fit for heterogeneous integration above integrated circuits. A 100 nm air gap excites the second-order symmetric Lamb wave (S2) mode in a 500 nm thick periodically poled piezoelectric film (P3F) LN from a fused silica substrate via flip-chip bonding, yielding a resonance frequency (\(f_s\)) of 7.16 GHz with an extracted \(k_t^2\) of 12.25\% and \(Q\) of 100, facilitating a high \(Q\cdot k_t^2\cdot f_s\) product of 87.71. The results demonstrate strong piezoelectric transduction through a nanoscale air gap, supporting further development of the platform for heterogeneous integration of compact, low-loss microwave components.
\end{abstract}


\section{Introduction}
\label{sec:introduction}
\IEEEPARstart{P}{iezoelectric} acoustic resonators are ubiquitously utilized as compact and low-loss frequency references for timing applications, enabling modern 5G communication technologies \cite{harpe_integrated_2023, nguyen_mems_2007, wu_mems_2020}. High-frequency bulk acoustic wave (BAW) devices in aluminum nitride (AlN) and scandium aluminum nitride (ScAlN) have been integral in realizing miniature oscillators fit for co-packaging or monolithic CMOS integration, presenting size advantages over well-established quartz crystals \cite{yen_high-frequency_2020,bahr_near-field-coupled_2021, bahr_transformer-coupled_2023, marigo_monolithic_2019}. However, these platforms have faced significant challenges in frequency scaling, inhibiting their adoption past 7 GHz for emerging 6G bands such as frequency range 3 (FR3, 7.125-24.25 GHz)~\cite{schaffer_33_2022, felmetsger_deposition_2011, andrews_6g_2024}.

To overcome these challenges, a laterally-excited bulk acoustic wave resonator (XBAR) platform has been developed in single-crystal lithium niobate (LN), leveraging its strong piezoelectricity to excite higher-order antisymmetric Lamb wave modes without the need for a bottom electrode. It has shown considerable promise for compact timing components with its ability to simultaneously realize high quality factor (\(Q\)) and electromechanical efficiency (\(k_t^2\)) with resonance frequencies (\(f_s\)) beyond 7 GHz, facilitating resonators with a high \(Q\cdot k_t^2 \cdot f_s\) product \cite{kourani_ku_2020,kourani_l_2021,chulukhadze_frequency_2023, chulukhadze_high-q_2025, kramer_thin-film_2023, kramer_acoustic_2025, lu_recent_2025}.

Despite the frequency scaling potential of the XBAR platform, its lateral-field-excited (LFE) nature has outlined new challenges for its adoption. LN is not a CMOS-compatible material, requiring external galvanic interconnects for system-level integration \cite{marigo_monolithic_2019}. Meanwhile, the XBAR has an inherent limitation in its \(C_0\)-density, creating susceptibility to parasitic reactive behavior posed by the interfacing circuitry \cite{chulukhadze_toward_2026}. Hence, the high stand-alone performance of LN XBARs is not yet accessible with state-of-the-art (SoA) circuit integration techniques, and motivates alternate resonator excitation schemes.

\begin{figure}[!t]
\centering
\includegraphics[width=\columnwidth]{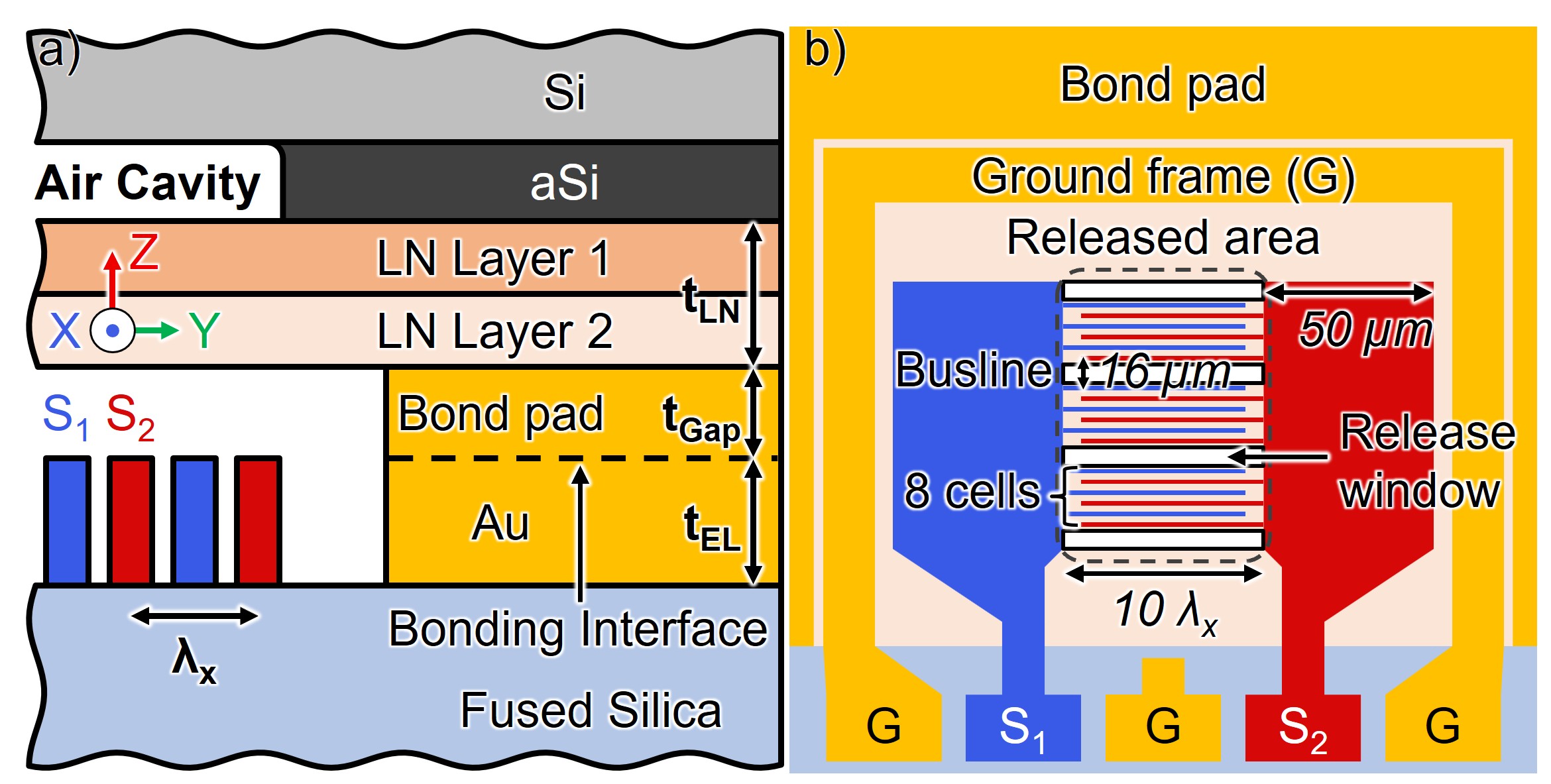}\\[-1pt]
\caption{(a) Cross-section and (b) top view of capacitively coupled XBAR with
key design parameters labeled.}
\label{fig:crosssectiontopview}
\end{figure}

This work presents a capacitively coupled implementation of a periodically poled piezoelectric film (P3F) LN XBAR which maintains the platform's high \(k_t^2\), alleviating the need for external interconnecting circuitry. Importantly, the device is fit for heterogeneous integration above IC, presenting a paradigm shift in high-frequency XBAR design for timing applications. We demonstrate an XBAR with cross-sectional and top views seen in Fig. \ref{fig:crosssectiontopview}(a) and (b), respectively, depicting a 500 nm thick P3F Z-cut LN capacitively coupled through a 100 nm thick air gap from a fused silica substrate, integrated using flip-chip bonding. The fabricated device yields a second-order symmetric Lamb wave (S2) mode resonance at 7.16 GHz with an extracted \(k_t^2\) of 12.25\% and \(Q\) of 100, facilitating a high \(Q\cdot k_t^2 \cdot f_s\) product of 87.71 with the initial prototype. Compared with prior capacitively excited piezoelectric resonators, the presented device obtains the highest reported \(k_t^2\), achieves a high resonance frequency, and shows promise for post-CMOS integration (Table \ref{tab:comparison}).

\section{Design and Simulation}
\label{sec:design}

The proposed platform relies on the dispersive nature of the laterally excited first-order antisymmetric Lamb wave (A1) mode in LN, facilitating considerable energy confinement \cite{lu_rf_2021, lu_5-ghz_2020}. This property enables capacitive excitation and localization of these waves in a piezoelectric solid possessing an appreciable relevant electromechanical coupling tensor component (\(e_\mathrm{15}\)). Prior works have leveraged this effect to demonstrate non-contact excitation of GHz LN resonators; however, these devices have suffered from a limited \(k_t^2\) and significant parasitic effects, which have made performance evaluation difficult \cite{wang_noncontact_2024, zheng_metal-air-piezoelectric_2025}. Here, we seek to retain a high \(k_t^2\) through a nominal 100-nm air gap with the transduction electrodes on the host substrate, eliminating galvanic interconnects to the LN membrane.

To target operation within FR3, the resonators were designed on a 500-nm-thick P3F Z-cut LN on an amorphous silicon (a-Si) on silicon (Si) stack, targeting the S2 mode resonance at 8 GHz~\cite{kramer_generalized_2025}. The wafer is provided by NGK Corporation. Z-cut LN is selected due to its high \(e_\mathrm{15}\) and its in-plane isotropy. The P3F film is utilized for enhanced film thickness while maintaining a high resonance frequency~\cite{kramer_trilayer_2023, zheng_periodically_2025}. We develop an architecture that simultaneously enables fabrication of a nm-scale air gap for capacitive transduction and has potential for heterogeneous integration post-CMOS, with a mock-up of an individual device cross-section seen in Fig.~\ref{fig:crosssectiontopview}~\cite{chulukhadze_device_2026}. The prototype relies on a gold-gold (Au-Au) bond in order to define the air gap using bond pads on the LN surface, which allows for thick Au electrodes on the substrate to minimize resistive loss. The design then aims to define an air cavity on the LN, which becomes the active area, followed by a flip-chip bonding process to define an individual device. A fused silica substrate is utilized due to the low RF loss at GHz frequencies and its optical transparency, enabling device inspection post-fabrication. Hence, the proposed method aims to create a die-to-chip framework toward heterogeneously integrated high-performance GHz acoustic resonators.

\begin{figure}[!t]
\centerline{\includegraphics[width=\columnwidth]{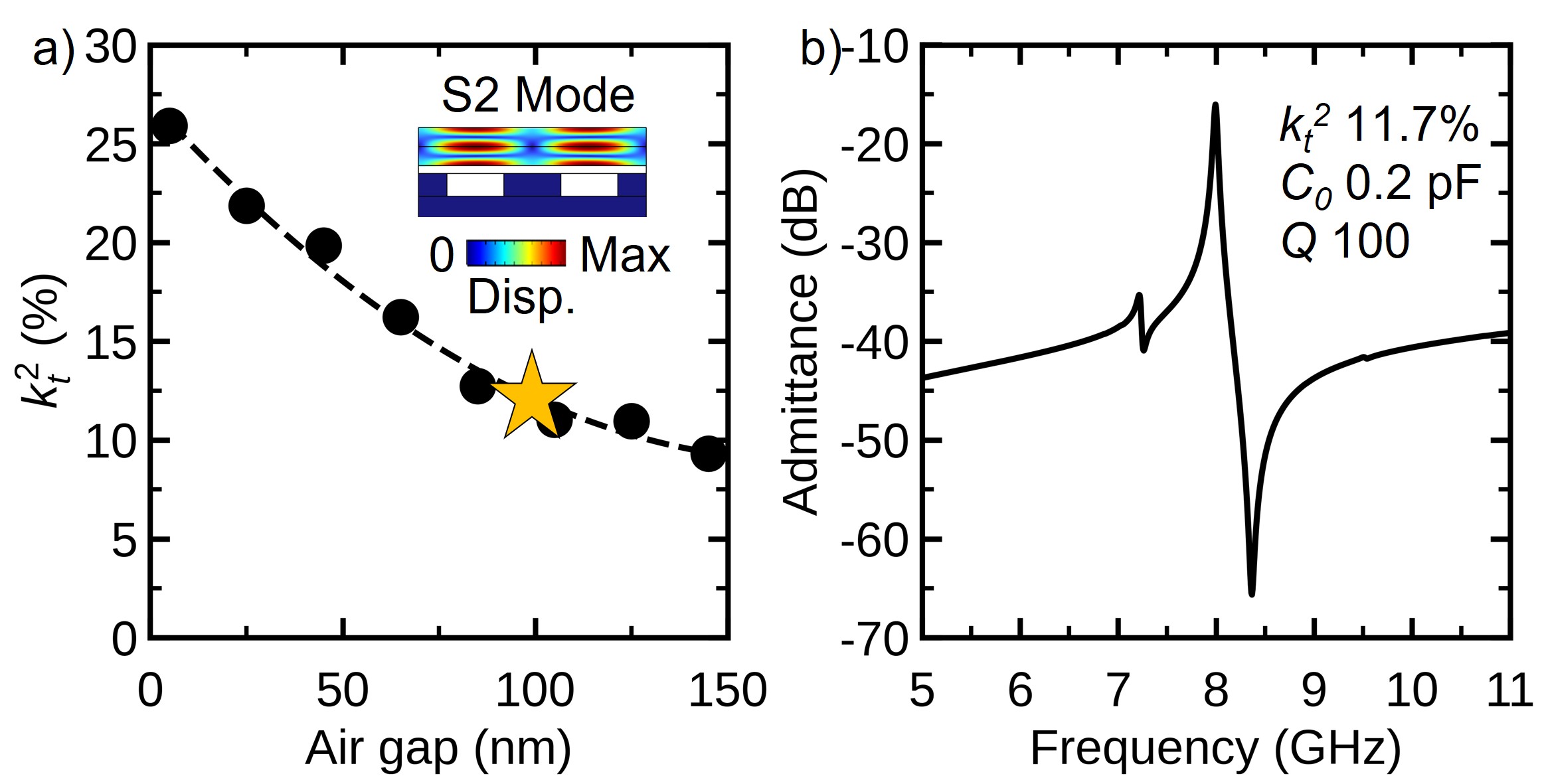}}
\caption{(a) Simulated \(k_t^2\)  for air gaps ranging from 10 to 150~nm with S2 mode-shape inset; (b) Simulated admittance for a 100 nm air gap.}
\label{fig:gapsim}
\end{figure}

\begin{figure}[!t]
\centerline{\includegraphics[width=0.95\columnwidth]{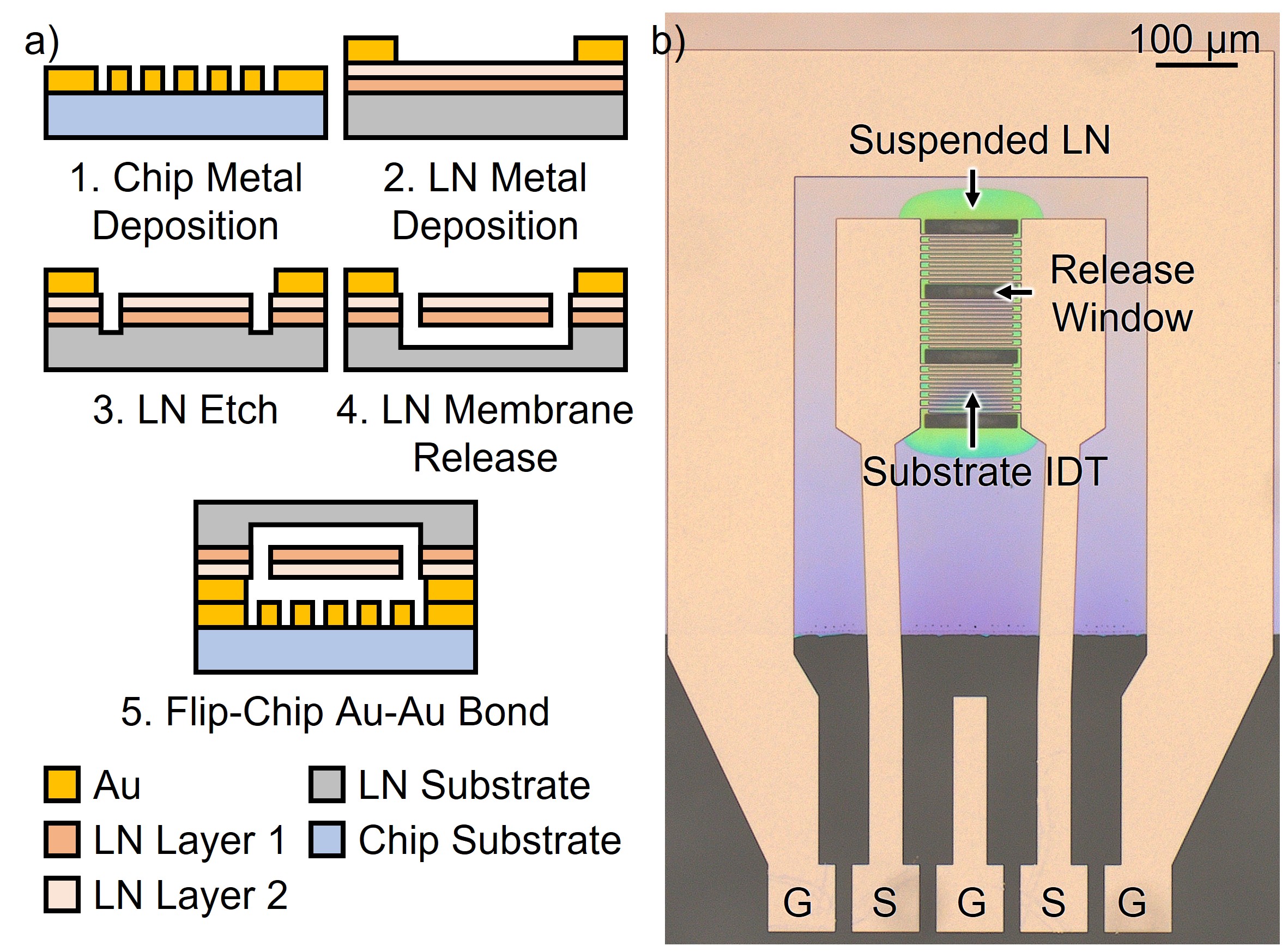}}
\caption{(a) Fabrication flow. (b) Optical image of the fabricated device taken through a transparent fused silica substrate.}
\label{fig:fab}
\end{figure}

The proposed capacitively coupled electrode geometry is analyzed using a two-dimensional unit-cell finite-element model in COMSOL, following \cite{chulukhadze_high-q_2025}. The air gap is swept from 10 to 150~nm to evaluate its effect on \(k_t^2\) [Fig.~\ref{fig:gapsim}(a)]. Increasing the gap reduces the electric field coupled into the LN film and consequently decreases \(k_t^2\). Compared with incumbent piezoelectric material platforms, substantial \(k_t^2\) is retained at the nominal 100-nm gap, which is selected as the design value.

The remaining design parameters comprise \(\lambda_x\) of the resonator, its transverse aperture, and the unit-cell layout. The criteria for their selection primarily concerned the structural stability of the membrane during the bonding process and aimed to balance device size with \(C_0\). A 10 \(\mu\)m \(\lambda_x\) is selected for fabrication simplicity, while the transverse aperture is selected as 10 \(\lambda_x\) to avoid significant transverse spurious modes. The unit cells were laid out with etch windows embedded every eight unit cells, enhancing the structural rigidity. In tandem, these design choices lead to the simulated performance seen in Fig. \ref{fig:gapsim}(b), revealing a high \(k_t^2\) of 11.7\% and a \(C_0\) of 0.2 pF with an assumed \(Q\) of 100.

\section{Fabrication and Measurement}
\label{sec:fab}

The resulting design was fabricated using standard cleanroom procedures. The process began with patterning and evaporation of Au on a 2\(\times\)2 cm\(^\text{2}\) fused silica substrate and a 7\(\times\)7 mm\(^\text{2}\) LN die at thicknesses of 300 nm and 100 nm, respectively. The LN chip was then etched with an ion milling tool in order to expose the sacrificial aSi layer for suspension while maintaining single-crystal quality \cite{chulukhadze_frequency_2023}. The membranes were suspended using an XeF\(_{2}\) gas etch, and the structure was thermocompression bonded using a flip-chip bonder. The fabrication flow and an optical image of the bonded device taken from the backside of the transparent fused silica substrate are shown in Fig. \ref{fig:fab}(a) and (b).

\begin{figure}[!t]
\centerline{\includegraphics[width=\columnwidth]{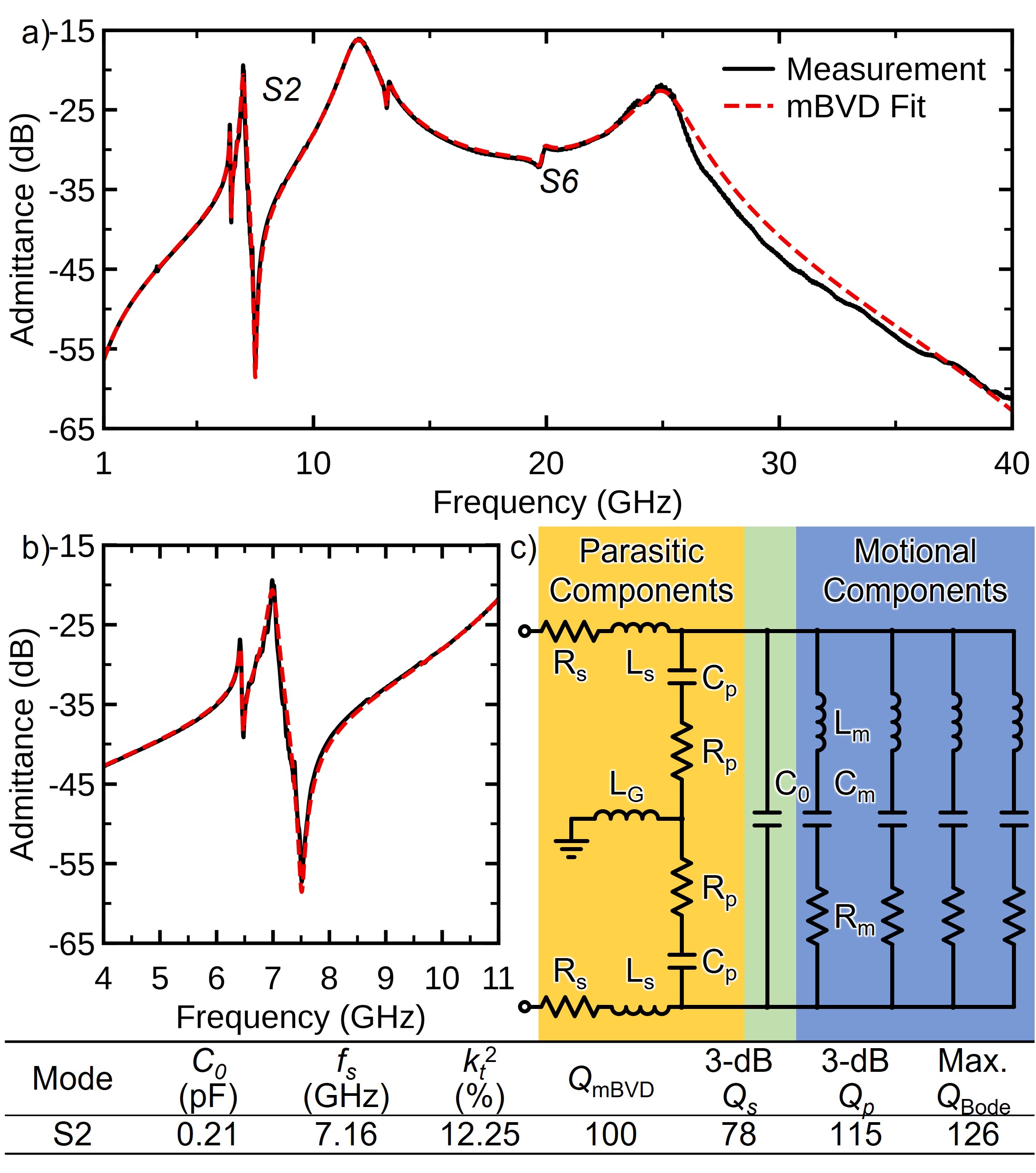}}
\caption{Measured (a) wideband and (b) narrowband admittance and its mBVD model fit. (c) mBVD model and key extracted parameters.}
\label{fig:meas}
\end{figure}

The devices were measured in a 2-port configuration using a GGB differential probe following standard calibration procedures. The measured admittance and its modified Butterworth-Van Dyke (mBVD) fit [Fig. \ref{fig:meas}(a)-(b)] reveal an S2 resonance at 7.16 GHz and an S6 resonance at 19.7 GHz, validating the proposed resonator geometry. Additional resonances near 12 and 25~GHz are attributed to electromagnetic (EM) parasitics associated with the routing geometry, which an above-IC implementation would not require. The mBVD model includes a series inductor (\(L_s\)) and resistor (\(R_s\)), a shunt capacitor (\(C_p\)) and resistor (\(R_p\)), and a ground inductor (\(L_g\)). The model closely fits the S2 and S6 modes, as well as all major spurious modes, in the presence of both EM resonances. The extracted key S2 parameters are summarized in the inset table in Fig. \ref{fig:meas}, revealing an extracted \(k_t^2\) of 12.25\%, an extracted \(Q\) of 100, and a \(C_0\) of 0.21 pF. The minor mismatch between the simulated and measured resonance frequencies, as well as \(k_t^2\), indicates a larger LN thickness than intended. Therefore, the measured performance parameters simultaneously validate the 100 nm air gap [Fig.~\ref{fig:gapsim}(a)] and the platform's ability to realize capacitively coupled high-\(k_t^2\) XBARs.

\begin{table}[!t]
\caption{Comparison to Prior Capacitive-Piezoelectric Resonators}
\label{tab:comparison}
\centering
\begingroup
\fontsize{9}{10.5}\selectfont
\setlength{\tabcolsep}{1pt}
\renewcommand{\arraystretch}{1.18}

\begin{tabular*}{\columnwidth}{
    @{\extracolsep{\fill}}
    M{20pt} M{33pt} M{25pt} M{29pt}
    M{24pt} M{33pt} M{32pt} M{31pt}
    @{}
}
\hline
Ref. & Material
& \shortstack{$f$\\(GHz)}
& $Q$
& \shortstack{$k_t^2$\\(\%)}
& \shortstack{\(Q{\cdot}k_t^2{\cdot}f_s\)\\(GHz)}
& \shortstack{$|Z_p/Z_s|$\\(dB)}
& \shortstack{Hetero.\\integ.} \\
\hline
\cite{besson_new_1977}
& Quartz & 0.005 & $3{\times}10^6$
& -- & -- & -- & N \\

\cite{xie_167-mhz_2020}
& AlN & 0.167 & 4536
& 0.358 & 2.71 & 44.9 & N \\

\cite{yen_high-q_2013}
& AlN & 0.941 & 5019
& 0.294 & 13.88 & 43.2 & N \\

\cite{hung_capacitive-piezoelectric_2015}
& AlN & 1.23 & 3073
& 1.12 & 42.35 & 57.9 & N \\

\cite{zheng_metal-air-piezoelectric_2025}
& Z LN & 10.38 & 188
& 2.85 & 55.62 & 26.4 & N \\

\cite{wang_noncontact_2024}
& Z LN & 6.72 & 1052
& 0.037 & 2.62 & 2.7 & Y \\
\hline
\textbf{This}
& \textbf{P3F LN}
& \textbf{7.16}
& \textbf{100}
& \textbf{12.25}
& \textbf{87.71}
& \textbf{40.1}
& \textbf{Y} \\
\hline
\end{tabular*}

\endgroup
\end{table}

Compared with prior capacitively coupled piezoelectric resonator demonstrations, shown in Table \ref{tab:comparison}, the proposed structure obtains the highest \(k_t^2\) alongside the highest \(Q\cdot k_t^2 \cdot f_s\) product and impedance ratio ($|Z_p/Z_s|$) beyond 7 GHz, and is fit for above-IC integration. Importantly, while the dispersive nature of excited Lamb waves ensured energy confinement, future studies will focus on further improving the \(Q\cdot k_t^2 \cdot f_s\) within FR3. With continued development, the presented XBAR platform holds strong potential as a heterogeneously integrated frequency-selective element for a timing source.

\section{Conclusion}
\label{sec:conclusion}
The presented capacitively coupled XBAR presents a paradigm shift in high-frequency resonator design, enabling non-contact, high-\(k_t^2\) resonators with potential for heterogeneous integration. We demonstrated an LN XBAR excited through a 100-nm air gap, with an S2 resonance at 7.16~GHz, an extracted $\kt$ of 12.25\%, and an extracted $Q$ of 100. Literature comparison shows that the proposed platform has overcome the primary limitation of the limited \(k_t^2\) in capacitive-piezoelectric devices, and its performance analysis yields a direct roadmap toward integration-ready, high-\(Q\cdot k_t^2 \cdot f_s\) heterogeneously-integrated acoustic resonators.

\clearpage
\bibliographystyle{IEEEtran}
\bibliography{IEEESettings,thebibliographycust}

\end{document}